\documentstyle[tighten,aps,prc,epsfig,floats]{revtex}
\begin{document}
\draft
\title{Total cross section of the
${}^{\bbox{3}}$H$(\bbox{p},\bbox{n}){}^{\bbox{3}}$He
reaction from threshold to 4.5~MeV}
\author{C.~R.~Brune\thanks{Present address:
  University of North Carolina at Chapel Hill,
  Chapel Hill, NC 27599-3255, USA and
  Triangle Universities Nuclear Laboratory,
  Durham, NC 27708-0308, USA},
  K.~I.~Hahn\thanks{Present address:
  Department of Science Education,
  Ewha Women's University, Seoul 120-750, Korea},
  R.~W.~Kavanagh, and
  P.~R.~Wrean\thanks{Present address: TRIUMF, Vancouver, BC, Canada~~V6T 2A3.}}
\address{W. K. Kellogg Radiation Laboratory, 106-38\\
California Institute of Technology\\
Pasadena, CA 91125, USA}
\date{\today}
\maketitle

\begin{abstract}
We report new measurements of the total cross section
for the ${}^3{\rm H}(p,n){}^3{\rm He}$ reaction from threshold
($E_p=1.02$~MeV) to $E_p=4.5$~MeV. The experiment utilized
specially prepared Ti--${}^3{\rm H}$ targets, and neutrons were
detected using a $4\pi$ detector.
A weak resonant structure due to an excited state in ${}^4{\rm He}$
is observed which was not seen in previous cross section measurements.
A new expression for the ${}^3{\rm He}(n,p){}^3{\rm H}$
thermonuclear reaction rate for temperatures below
10~GK is presented which will allow for
more accurate calculations of
the yields of light elements produced by big-bang nucleosynthesis.
 
\end{abstract}


\section{Introduction}

The ${}^3{\rm H}(p,n){}^3{\rm He}$ reaction and its inverse
${}^3{\rm He}(n,p){}^3{\rm H}$ are important in many subfields of physics.
Due to its large cross section and other properties,
this reaction is commonly used for two purposes in neutron physics:
the ${}^3{\rm H}(p,n){}^3{\rm He}$ reaction is an important source of
neutrons, and the ${}^3{\rm He}(n,p){}^3{\rm H}$ reaction
is often used for detecting neutrons.
This reaction also provides information about the excited levels
of ${}^4{\rm He}$, which are still not well understood~\cite{Hof97,Cos97}.
The ${}^3{\rm H}(p,n){}^3{\rm He}$ reaction near threshold
is strongly influenced by the first two excited levels of ${}^4{\rm He}$,
which lie 370 keV below and 430 keV above the
${}^3{\rm H}(p,n){}^3{\rm He}$ threshold, respectively~\cite{Til92}.
The present experiment is primarily motivated by
the role of this reaction in big-bang nucleosynthesis.

The standard big-bang model of the primordial universe is very successful in
accounting for the observed relative abundances of the light elements
${}^2{\rm H}$, ${}^3{\rm He}$, ${}^4{\rm He}$, and ${}^7{\rm Li}$
\cite{Kra90,Ril91,Wal91,Smi93,Cop95}. The calculated abundances
agree with observations only for baryon densities significantly lower
than the critical density, in the range
$0.01\lesssim\Omega_B\lesssim0.1$.
The uncertainties in the abundance calculations arising from nuclear-data
input have been studied in detail by Krauss and Romanelli~\cite{Kra90}
and Smith, Kawano, and Malaney~\cite{Smi93}.
The latter have identified 12 reactions which significantly
affect light-element productions. Their assumed
${}^3{\rm He}(n,p){}^3{\rm H}$ reaction rate was found to have a significant
impact on the calculated abundances of ${}^3{\rm He}$ and ${}^7{\rm Li}$.
At the temperatures important for determining big-bang yields, the
reaction rate is determined by the ${}^3{\rm He}(n,p){}^3{\rm H}$
cross section in the energy range $1 \lesssim E_{c.m.} \lesssim 250$~keV
(see Sec.~\ref{sec:nasv} below).
In this energy range, the existing data are sparse, and not in good
agreement.

Previous measurements of the ${}^3{\rm H}(p,n){}^3{\rm He}$ total cross
section for $E_p\le 4.5$~MeV have been reported by
Vlasov {\em et al.}~\cite{Vla55}, Gibbons and Macklin~\cite{Gib59},
Perry {\em et al.}~\cite{Per59}, and Macklin and Gibbons~\cite{Mac66}.
Unfortunately, two of the most important
total cross section measurements~\cite{Per59,Mac66}
have been published only in conference proceedings or as a laboratory report.

The ${}^3{\rm He}(n,p){}^3{\rm H}$
cross section at thermal energies
has been determined with high precision by the total cross section
of Als-Nielsen and Dietrich~\cite{Als64}.
Measurements at higher neutron energies have been reported by
Coon~\cite{Coo50}, Batchelor, Aves, and Skyrme~\cite{Bat55},
Sayres, Jones, and Wu~\cite{Say61},
Costello, Friesenhahn, and Lopez~\cite{Cos70}, and
Borzakov {\em et al.}~\cite{Bor82}.
Ratios of the ${}^3{\rm He}(n,p){}^3{\rm H}$ cross section
to the ${}^6{\rm Li}(n,\alpha)$ and
${}^{10}{\rm B}(n,\alpha)$ cross sections have been measured
by Bergman and Shapiro~\cite{Ber61} ($E_n\le 30$~keV)
and Bowman {\em et al.}~\cite{Bow80} ($E_n\le 25$~keV), respectively.

Recommended cross sections based on the available experimental data
have been given by Costello~\cite{Cos71}, Liskien and Paulsen~\cite{Lis73},
Drosg~\cite{Dro80}, Drosg {\em et al.}~\cite{Dro85},
B{\" o}dy~\cite{Bod87}, Drosg and Schwerer~\cite{Dro87},
Hale, Dodder, and Young (ENDF/B-VI)~\cite{endf},
Smith, Kawano, and Malaney~\cite{Smi93}, and Drosg~\cite{Dro96}.
The ${}^3{\rm H}(p,n){}^3{\rm He}$ cross section at threshold is
determined to better than $1\%$ from the thermal
${}^3{\rm He}(n,p){}^3{\rm H}$ cross section~\cite{Als64},
and the standard deviation in the evaluated cross section is estimated
to be $4\%$ or less for $2 \le E_p \le 16$~MeV~\cite{Dro87}.
Between threshold and 2~MeV,
most of the experiments have errors of $10\%$ or greater.
The recent and accurate (2-3\% quoted uncertainty) measurements
of the ${}^3{\rm He}(n,p){}^3{\rm H}$ cross section
for $E_n \le 137$~keV~\cite{Bor82} disagree with the energy dependence found
by Ref.~\cite{Mac66} by about 25\%, and also
disagree with the ENDF/B-VI evaluation by up to 15\%.
The energy range for big-bang nucleosynthesis lies
between threshold and 2~MeV; it is the large uncertainty within
this energy range which motivates the present work.

In this paper, we report new measurements of the
${}^3{\rm H}(p,n){}^3{\rm He}$ total cross section,
from threshold to $E_p=4.5$~MeV.
A new thermonuclear reaction rate is calculated
which is valid for temperatures less than 10~GK.

\section{Experimental Apparatus and Procedures}
\label{sec:exp}

\subsection{Beam}
\label{subsec:beam}

The proton, deuteron, and ${}^4{\rm He}^+$ beams used in this
experiment were supplied by the Caltech 3-MV Pelletron Tandem Accelerator.
The beam energy was defined by a $90^\circ$
analyzer magnet and NMR magnetometer.
The energy calibration ($\pm 0.1\%$) was established using the resonances
at  $483.91\pm0.10$-keV \cite{Ajz87}
in ${}^{19}{\rm F}(p,\alpha\gamma)$,
$991.86\pm0.03$-keV \cite{End90} in ${}^{27}{\rm Al}(p,\gamma)$,
$606.0\pm0.5$-keV \cite{Wan91} in ${}^{11}{\rm B}(\alpha,n)$,
and $1530.03\pm0.15$-keV \cite{Maa78} in
${}^{24}{\rm Mg}(\alpha,\gamma)$.
A  collimator limited the beam to an area of $\approx 1$~cm$^2$
on the targets; the beam was rastered over this area in order to produce
a uniform intensity distribution.
Beam currents for the ${}^3{\rm H}(p,n){}^3{\rm He}$ measurements
varied between 7 and 100~nA, depending on the desired counting rate.
The number of incident particles was determined by beam-current
integration, to a precision of $\pm 1\%$.

\subsection{Targets}
\label{subsec:targets}

The preparation, characterization, and use of the Ti--${}^3{\rm H}$
targets have been described previously \cite{Bru94a,Bru94b,Bru94c}.
Briefly, Ti was evaporated onto
31.7-mm-diameter, 0.81-mm-thickness Cu or Ta substrates; the substrates were
then heated in an atmosphere of ${}^3{\rm H}_2$ gas to induce the
formation of Ti--${}^3{\rm H}$. The substrates were maintained
in high vacuum during the time between Ti evaporation and tritide formation,
minimizing contamination and maximizing the attainable ${}^3{\rm H}$:Ti
ratio.

The ${}^3{\rm H}$ and Ti areal densities were determined 
in a scattering chamber using the
${}^3{\rm H}(d,\alpha)$ and ${\rm Ti}(\alpha,\alpha)$ reactions, respectively.
The total cross section and center-of-mass Legendre coefficients for
${}^3{\rm H}(d,\alpha)$ reaction, needed for the absolute ${}^3{\rm H}$
areal-density determination, were taken from the evaluation of Drosg and
Schwerer \cite{Dro87}; the uncertainty in the cross section is estimated
to be $1.5\%$ for $E_d<400$~keV, increasing to $4\%$ for higher energies.
The absolute uncertainty in the areal density determinations is estimated
to be $\pm 4\%$.

Two targets were utilized for the ${}^3{\rm H}(p,n){}^3{\rm He}$ measurements.
One target consisted of $5.52\times 10^{17}$ and $2.96\times 10^{17}$
atoms/cm$^2$ of ${}^3{\rm H}$ and Ti, respectively, on a Cu backing.
This target is referred to as ``target 1'' in Ref.~\cite {Bru94c};
${\rm Ti}(\alpha,\alpha)$ and ${}^3{\rm H}(d,\alpha)$ spectra
obtained with this target are shown in Figs.~4 and 6 of Ref.~\cite{Bru94a}.
The use of this target for ${}^3{\rm H}(p,n){}^3{\rm He}$ measurements
is limited to energies below the ${}^{65}{\rm Cu}(p,n)$ threshold
at $E_p=2.17$~MeV.
To facilitate measurements at higher energies an additional target
was fabricated on a Ta backing with
$8.30\times 10^{17}$ and $9.5\times 10^{17}$
atoms/cm$^2$ of ${}^3{\rm H}$ and Ti, respectively.
The ${}^3{\rm H}(d,\alpha)$ excitation function obtained with this
target is shown in Fig.~7 of Ref.~\cite{Bru94a}.
The ${}^3{\rm H}(p,n){}^3{\rm He}$ measurements reported here were
performed before these targets were used for any other experiments.

Two additional targets were used to investigate backgrounds from
targets containing no ${}^3{\rm H}$.
One target consisted of $4.36\times 10^{17}$ and $2.98\times 10^{17}$
atoms/cm$^2$ of ${}^2{\rm H}$ and Ti, respectively, and was
produced on a Cu backing using the same equipment
as for the ${}^3{\rm H}$ targets.
The other target consisted of $1.0\times 10^{18}$ Ti atoms/cm$^2$
evaporated on a Ta backing, and was not hydrided.

\subsection{Neutron detection}
\label{subsec:neutrons}

Neutrons were detected in a $4\pi$ detector consisting of up to 12
${}^3{\rm He}$-filled proportional counters embedded in a polyethylene
moderator surrounding the target chamber.
The efficiency is slowly varying in the region $0.0005 < E_n < 2$~MeV,
such that it can be approximated by a constant within $\pm 10\%$.
However, the efficiency drops dramatically outside of this range,
by about a factor of two for 1-eV or 5-MeV neutrons.
For a known distribution of neutron energies and emission angles,
the efficiency can be more accurately determined from Monte Carlo simulations.
For this purpose we utilize the computer code {\sc mcnp}~\cite{mcnp}, which
simulates the transport and detection of neutrons, given the
materials and geometry of our target holders and detector.
Additional information concerning the detector, the Monte Carlo simulations,
and experimental validation of the simulations
is given in Refs.~\cite{Wre98,Wre99}.

\begin{figure}
\begin{center}
\includegraphics[bbllx=43,bblly=64,bburx=532,bbury=701,width=3.4in]{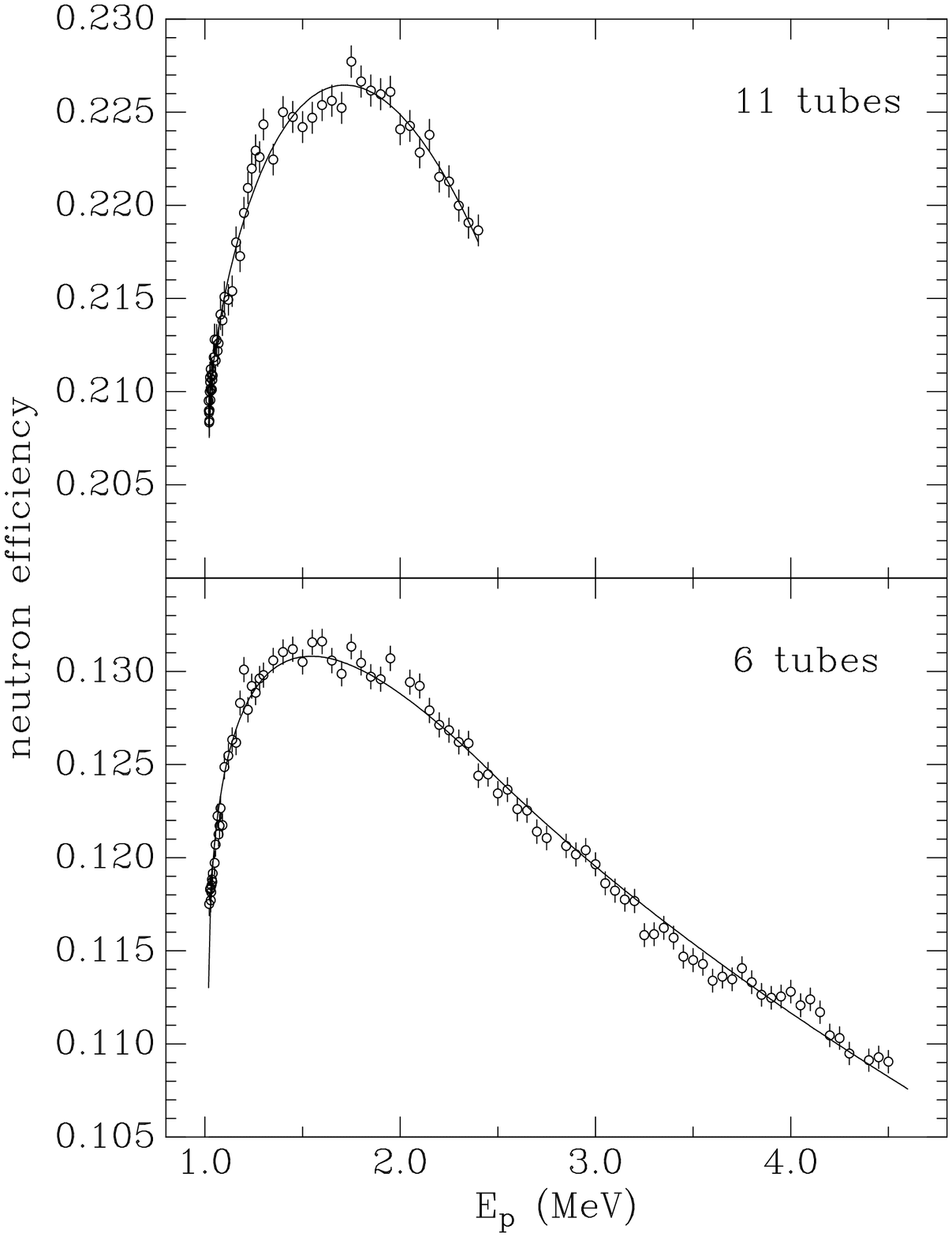}
\end{center}
\caption{The neutron efficiency as a function of proton energy for
the 11-tube configuration (top panel) and the 6-tube
configuration (bottom panel). The Monte Carlo calculations
with statistical errors are shown
as circles, and the solid curves are empirical fits used in subsequent
calculations. Note the break in vertical scale between the two panels.}
\label{fig:effpn}
\end{figure}

One set of ${}^3{\rm H}(p,n){}^3{\rm He}$
measurements was taken using 11 detection tubes in the moderator,
with one additional tube present but not used.
The other set of measurements was taken using 6 tubes in the moderator.
The angular distributions of neutrons emitted from the
${}^3{\rm H}(p,n){}^3{\rm He}$ reaction, required to determine the
(correlated) distributions of neutron energies and emission angles
in the Monte Carlo simulations, were taken from Ref.~\cite{Dro87}.
In order to investigate the sensitivity of the simulations to the
assumed angular distributions, we also performed simulations assuming
that the reaction is isotropic in the center-of-mass system.
The resulting efficiencies differed by at most 2\% for $E_p\le 4.5$~MeV,
indicating an uncertainty of $<1\%$ from the assumed angular distributions.
The simulated efficiencies for the 6-tube configuration
were renormalized by 1.02, so that the ratio of measured
to simulated efficiency for a ${}^{252}{\rm Cf}$ neutron source was
the same for both sets of measurements.
This procedure standardizes the measurements to the configuration
for which the efficiency of the detector has been
extensively tested~\cite{Wre98}.
The results of the simulations for the
two detector configurations are shown in Fig.~\ref{fig:effpn}.
Also shown as solid lines are empirical fits used in subsequent analysis.
We estimate the systematic uncertainty in the neutron detection
efficiency to be $\pm 3\%$.

\subsection{Yield measurements}
\label{subsec:yield}

The neutron yield with the Cu-backed ${}^3{\rm H}$ target perpendicular
to the beam was measured with the 11-tube detector configuration
for $1.016 \le E_p \le 2.15$~MeV, i.e.,
from just below the ${}^3{\rm H}(p,n){}^3{\rm He}$ threshold to
just below the ${}^{65}{\rm Cu}(p,n)$ threshold.

\begin{figure}
\begin{center}
\includegraphics[bbllx=51,bblly=64,bburx=532,bbury=701,width=3.4in]{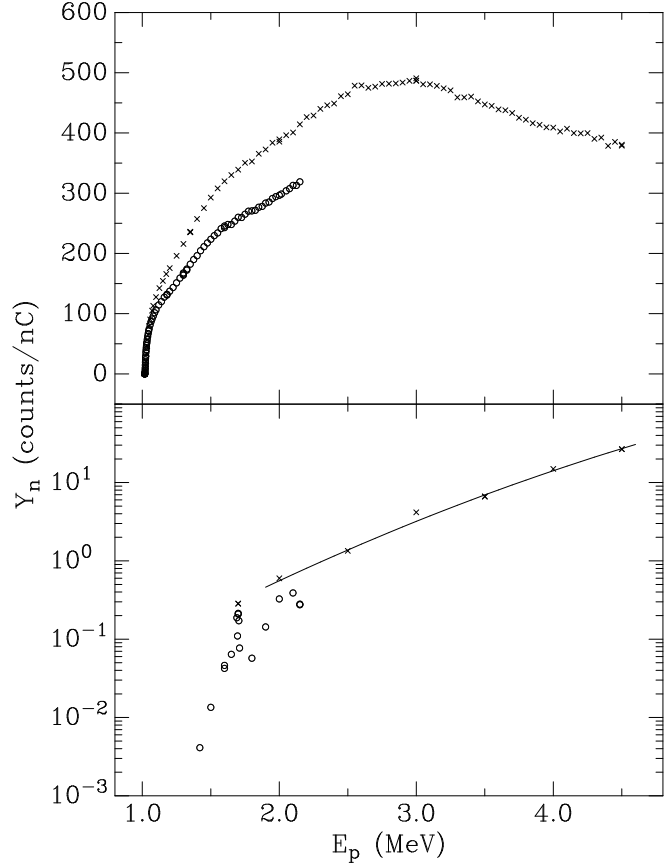}
\end{center}
\caption{Yields $Y_n$ of detected neutrons for various targets.
Upper panel: For the Cu-backed ($\circ$) and Ta-backed ($\times$)
${}^3{\rm H}$ targets.
Lower panel: For the Cu-backed ($\circ$) and Ta-backed ($\times$)
targets used for background measurements.
The statistical errors are smaller than the size of the data points.
The solid curve is an empirical fit used for background subtraction.}
\label{fig:yield}
\end{figure}

The yield with the Ta-backed ${}^3{\rm H}$ target at $45^\circ$ with
respect to the beam was measured with 6~tubes
from just below threshold to $E_p=4.5$~MeV.
The measured yields are displayed in the upper panel of Fig.~\ref{fig:yield}.
The yields were corrected for dead time in the detector
($\le 2\%$ for all measurements).
Repeated measurements at a standard energy of $E_p=1.3$~MeV indicated that
the target thickness deteriorated by less than 1\%
over the course of the measurements.
The stability of the detection system was monitored
throughout the experiment by measuring the efficiency for a
${}^{252}{\rm Cf}$ neutron source.

Beam-off backgrounds were completely negligible for this experiment.
It is however important to consider possible backgrounds from other
neutron-producing reactions which may take place in the target.
Of particular concern here are the ${}^{49}{\rm Ti}(p,n)$
and ${}^{50}{\rm Ti}(p,n)$ reactions which have thresholds
at proton energies of 1.41 and 3.05~MeV, respectively.
In order to test for possible backgrounds, measurements were made
using targets without ${}^3{\rm H}$
(these targets are described in Subsec.~\ref{subsec:targets}).
The Ti--${}^2{\rm H}$ target was measured with the same target angle and
neutron detector configuration as the Cu-backed ${}^3{\rm H}$ target;
the Ti target on Ta backing was measured under the same conditions as
the Ta-backed ${}^3{\rm H}$ target. The results of these
measurements are shown on the lower panel of Fig.~\ref{fig:yield}.

\section{Data Analysis and Results}
\label{sec:results}

The magnitude and energy dependence of the beam-dependent
background observed from targets containing no ${}^3{\rm H}$
are consistent with the known ${}^{48}{\rm Ti}(p,n)$~\cite{Ken80a} and
${}^{49}{\rm Ti}(p,n)$~\cite{Ken80b} cross sections.
Using the known Ti areal density, the background present in the
measurements using the Cu-backed ${}^3{\rm H}$ target is
estimated to be at most 0.2\%, and was therefore neglected.
For the Ta-backed ${}^3{\rm H}$ target, the background is estimated
to be $<1\%$ for $E\le 3.2$~MeV.
For $E_p\ge 2.0$~MeV the yields measured from this target were
corrected for background, using the solid curve shown in Fig.~\ref{fig:yield},
scaled by the ratio of Ti areal densities.
This correction was at most 7\%, at $E_p=4.5$~MeV.
An error of $\pm 30\%$ was assumed for the subtracted yield.

The background-corrected yield of neutrons detected per incident particle
$Y_n$, for a mono-energetic beam of energy $E_p$,
is given by
\begin{equation}
  Y_n = (nt)~\sigma(E_p)~\varepsilon(E_p)~, 
  \label{eq:yield}  
\end{equation}
where $(nt)$ is the areal number density of target atoms, $\sigma$ is
the ${}^3{\rm H}(p,n){}^3{\rm He}$ cross section, and $\varepsilon$ is the
neutron-detection efficiency.  For a target of significant thickness,
the beam loses energy as it passes through the target, and the
yield is then given by
\begin{equation}
  Y_n = {\int_{E_{\rm t}}^{E_p}
  \sigma (E_p^{'})~\varepsilon(E_p^{'}) 
  \biggl[ {{{dE}\over{dX}} (E_p^{'})} \biggr]^{-1} dE_p^{'}}~,
  \label{eq:conv}
\end{equation}
in which $E_{\rm t}=E_p-\Delta E$,
where $E_{p}$ is the incident proton energy and
$\Delta E$ is the energy loss in the target.
The energy loss of protons in the target per ${}^3{\rm H}$ atom
per unit area is given by
\begin{equation}
  {{dE}\over{dX}} = \biggl({{dE}\over{dX}}\biggr)_{\rm H}
  +{1\over r} \biggl({{dE}\over{dX}}\biggr)_{\rm Ti}~,
\end{equation}
where $({{dE}\over{dX}})_{\rm H}$ and
$({{dE}\over{dX}})_{\rm Ti}$ are the stopping powers for protons
in H and Ti~\cite{Zie77}, and $r$ is the ${}^3{\rm H}$:Ti ratio
which is assumed to be independent of depth in the target.
The energy loss for 2-MeV $\alpha$ particles in the Ti--${}^3{\rm H}$
computed by this method has been verified experimentally
for these targets~\cite{Bru94a}.

\begin{figure}
\begin{center}
\includegraphics[bbllx=35,bblly=38,bburx=522,bbury=724,%
angle=90,width=3.4in]{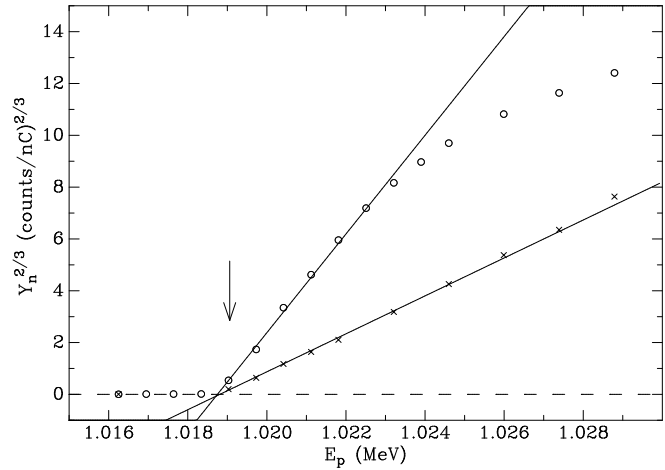}
\end{center}
\caption{The near-threshold yield of detected neutrons
(raised to the 2/3 power) for the
Cu-backed target ($\circ$) and the Ta-backed target ($\times$).
The statistical errors are smaller than the size of the data points.
The solid lines are linear fits described in the text.
The arrow indicates the known threshold energy.
The different slopes for the two targets are caused by differences in
neutron-detection efficiency and the ${}^3{\rm H}$:Ti ratio.}
\label{fig:threshold}
\end{figure}

Some indication of the consistency of the beam-energy calibration and
the quality of the targets is provided by the behavior of the neutron
yield near the threshold.
For a target sufficiently thick to integrate the cross section
down to the threshold energy $E_0$, the detected neutron yield
varies as $Y_n\propto (E_p-E_0)^{3/2}$, provided the neutron production
cross section varies as $(E_p-E_0)^{1/2}$
(i.e., assuming $s$-wave neutrons and no narrow resonance).
This simple analysis also assumes that the energy dependences
of the stopping power and detection efficiency are negligible.
A linear fit to $Y_n^{2/3}$ should thus intersect the $E_p$ axis at
the threshold energy.
Plots of $Y_n^{2/3}$ versus $E_p$ for the two targets are shown in
Fig.~\ref{fig:threshold}.
The yield from the Cu-backed target departs from a linear dependence
more quickly than the Ta-backed target due to its smaller thickness.
Linear fits are also shown which included points within $\approx 4$
keV of threshold for the Cu-backed target and within $\approx 10$~keV of
threshold for the Ta-backed target.
Note that 1-MeV protons lose 4.8~keV and 22~keV in the Ti--${}^3{\rm H}$
layer for the Cu-backed and Ta-backed targets, respectively.
Both fits give a threshold energy of 1.0188~MeV, which is in excellent
agreement with the known value 1.01906~MeV\footnote{All thresholds and
detailed balance conversions in this paper are computed using relativistic
kinematics with nuclear masses; $Q$-values
and atomic masses are taken from Audi and Wapstra~\protect\cite{Aud95}.
The uncertainty in the $Q$ value ($\approx 0.002$~keV) is sufficiently small to
be a negligible consideration in the detailed-balance conversions.}.

Cross sections were extracted from the measured yields using
Eqs.~(7)-(9) of Ref.~\cite{Wre94} to correct for the energy loss
effects described by Eq.~(\ref{eq:conv}).
This procedure requires that the energy dependence
of the cross section be known in advance.
For this purpose we assumed the ENDF/B-VI evaluation, converted to
${}^3{\rm H}(p,n){}^3{\rm He}$ using detailed balance.
The assumed energy dependence is only important
near threshold where the cross section changes significantly
as the beam loses energy in the target.
For $E_p\ge 1.1$~MeV this procedure
differs negligibly from using Eq.~(\ref{eq:yield}) with $E_p$
replaced by $E_p-\Delta E/2$, due to the thin targets used in the experiment.
The data were also analyzed assuming $\sigma\propto (E_p-E_0)^{1/2}$,
which for all incident energies
changed the resulting cross sections by an amount negligible
compared to other uncertainties.
One important source of error near the threshold is
the energy of the incident proton beam.
Given the good agreement obtained with the known threshold energy
already described, we have allowed for an
uncertainty of $\pm 0.5$~keV in the incident energy.
We have also allowed for $\pm 10\%$ uncertainty in the proton energy loss.
We do not present the data where $E_p \le E_0+\Delta E$ due to the
large errors from uncertainties in the incident energy and energy loss.
The error bars on the data points include uncertainties
from counting statistics, incident energy, energy loss,
and background subtraction.
Additional systematic errors in the data
are summarized in Table~\ref{tab:errors}.

The absolute cross sections determined from the Ta-backed target
are $\approx 2\%$ higher than from the Cu-backed target.
The data sets were renormalized to a scale corresponding to
the arithmetic mean of the two determinations.
The final results for the total cross section are shown in
Fig.~\ref{fig:sig_tot}.
The behavior of the cross section near the threshold is more
easily seen by converting the data to ${}^3{\rm He}(n,p){}^3{\rm H}$
cross sections using detailed balance and multiplying by $E_n^{1/2}$,
as shown for the near-threshold data in Fig.~\ref{fig:snl}.
The rather large systematic errors associated with the lowest-energy
data points shown in Fig.~\ref{fig:snl} arise from
the uncertainty in proton energy, as the detailed-balance conversion
is very sensitive to the proton energy when the proton energy is
near threshold.

\section{Discussion}
\label{sec:dis}

\subsection{Comparison to previous measurements}

We  will not attempt to compare our new results to all of the
data available with $E_p\le 4.5$~MeV. The reader is referred to the
evaluations~\cite{Smi93,Cos71,Lis73,Dro85,Bod87,Dro87,endf}, and
in particular Ref.~\cite{Dro80} for a critique of the previous measurements.
Some additional information related to previous experiments is
also supplied in the Appendix.
In Fig.~\ref{fig:sig_tot} our cross-section data are compared to two
recent evaluations~\cite{endf,Dro96}.
It is seen that both evaluations are in excellent agreement with the
present data, with a maximum deviation of $\approx 5$\% near
$E_p=2.7$~MeV, but within our estimated systematic error.

In Figs.~\ref{fig:snl_comp1} and~\ref{fig:snl_comp2} the data
are compared to some of the previous data in the energy range
appropriate for big-bang nucleosynthesis.
The data are in general agreement with most of the other older measurements
which had considerably larger errors.
The data agree well with results of Borzakov {\em et al.}~\cite{Bor82},
except for their higher-energy data points.
Our results are not in agreement with the data from
Macklin and Gibbons~\cite{Mac66}, especially for their higher energies.
It should be noted however that the scale of their data
is not absolute, so the discrepancy is in the energy dependence.

\subsection{Excited levels of ${}^4{\rm He}$}

The present data show the near-threshold energy dependence of the cross
section much more clearly than previous measurements.
In particular, it is clear from inspection of Fig.~\ref{fig:snl} that
the cross section deviates from the $1/v$ energy dependence
expected for non-resonant $s$-wave neutron-induced reactions
(a $1/v$ dependence would yield a
horizontal line when multiplied by $E_n^{1/2}$).
The observed energy dependence results primarily from the first three
excited levels of ${}^4{\rm He}$ located~\cite{Til92} at
$E_x=20.21$~MeV ($J^\pi=0^+$), 21.01~MeV ($J^\pi=0^-$), and
21.84~MeV ($J^\pi=2^-$); note that
the location of the ${}^3{\rm He}+n$ threshold is at $E_x=20.58$~MeV.
As discussed by Borzakov {\em et al.}~\cite{Bor82} and references therein,
the cross section very near threshold is dominated by the
subthreshold $0^+$ $s$-wave resonance.

\begin{figure}
\begin{center}
\includegraphics[bbllx=35,bblly=25,bburx=530,bbury=732,%
angle=90,width=3.4in]{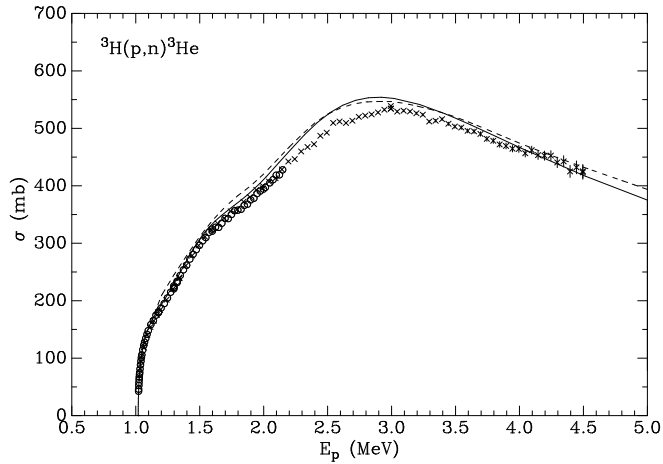}
\end{center}
\caption{The present results for the ${}^3{\rm H}(p,n){}^3{\rm He}$
total cross section, for the Cu-backed target
($\circ$) and Ta-backed target ($\times$).
When not visible, the errors are smaller than the size of the data points
(the additional systematic uncertainty of 5\% described in
Table~\protect\ref{tab:errors} is not included).
The solid curve is ENDF/B-VI evaluation~\protect\cite{endf} and
the dashed curve is the evaluation from the
{\sc drosg-96} computer code~\protect\cite{Dro96}.} 
\label{fig:sig_tot}
\end{figure}

\begin{figure}
\begin{center}
\includegraphics[bbllx=38,bblly=23,bburx=530,bbury=742,%
angle=90,width=3.4in]{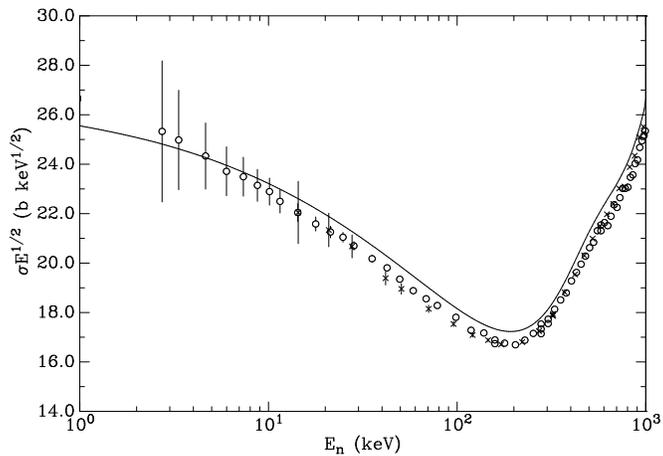}
\end{center}
\caption{The results, converted to ${}^3{\rm He}(n,p){}^3{\rm H}$,
and multiplied by $E_n^{1/2}$. The plot symbols have the same meaning
as in Fig.~\protect\ref{fig:sig_tot}.
When not visible, the errors are smaller than the size of the data points
(the systematic uncertainty of 5\% described in
Table~\protect\ref{tab:errors} is not included).
The solid curve is the ENDF/B-VI evaluation~\protect\cite{endf}.}
\label{fig:snl}
\end{figure}

\begin{figure}
\begin{center}
\includegraphics[bbllx=38,bblly=23,bburx=530,bbury=742,%
angle=90,width=3.4in]{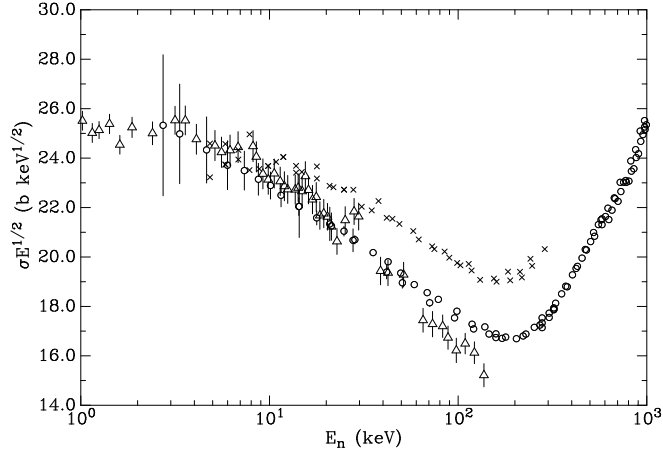}
\end{center}
\caption{The present results ($\circ$), compared to the data of
Macklin and Gibbons~\protect\cite{Mac66} ($\times$) and the data of
Borzakov~{\em et al.}\protect\cite{Bor82} ($\protect\bigtriangleup$).
All data are converted to ${}^3{\rm He}(n,p){}^3{\rm H}$ cross sections
and multiplied by $E_n^{1/2}$.}
\label{fig:snl_comp1}
\end{figure}

\begin{figure}
\begin{center}
\includegraphics[bbllx=38,bblly=16,bburx=530,bbury=742,%
angle=90,width=3.4in]{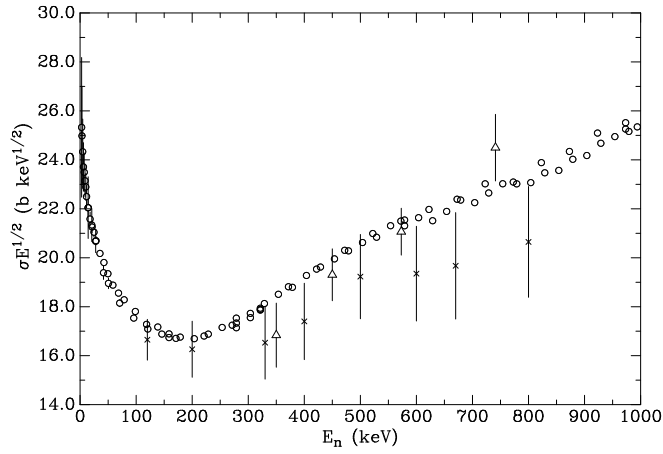}
\end{center}
\caption{The present results ($\circ$), compared to the data of
Batchelor, Aves, and Skyrme~\protect\cite{Bat55} ($\times$)
and the data of Costello, Friesenhahn, and
Lopez~\protect\cite{Cos70} ($\protect\bigtriangleup$).
All data are converted to ${}^3{\rm He}(n,p){}^3{\rm H}$ cross sections
and multiplied by $E_n^{1/2}$.}
\label{fig:snl_comp2}
\end{figure}

At higher energies, the effects of the other levels become important.
The present data as shown in Fig.~\ref{fig:sig_tot} indicate
a definite change in curvature near $E_p=1.6$~MeV.
In previous experiments this feature was masked by the larger errors
and/or coarser energy steps.
Interestingly, this feature is predicted very well by the
ENDF/B-VI evaluation~\cite{endf}
(see Fig.~\ref{fig:sig_tot} of the present work).
The ENDF/B-VI evaluation for ${}^3{\rm He}(n,p){}^3{\rm H}$ is
generated from an $R$-matrix analysis~\cite{Hal97} which is essentially
the same as that described in Ref.~\cite{Til92}.
This charge-independent $R$-matrix analysis includes cross-section
and polarization data for $n-{}^3{\rm He}$, $p-{}^3{\rm H}$,
and ${}^2{\rm H}-{}^2{\rm H}$ scattering and reactions;
and $p-{}^3{\rm He}$ and $n-{}^3{\rm H}$ scattering.
The curvature near $E_p=1.6$~MeV results from the $0^-$ second excited
state of ${}^4{\rm He}$. A recent measurement of the longitudinal
polarization-transfer coefficient near $E_p=1.6$~MeV~\cite{Wal98}
has provided striking evidence for this level. These authors conclude
that at the peak of the $0^-$ resonance near $E_p=1.6$~MeV, the reaction
is dominated by $0^+$ and $0^-$ amplitudes
which are approximately equal in strength.
At higher energies additional levels become important, especially
the $2^-$ state at $E_x=21.84$~MeV which gives rise to the broad peak
in the ${}^3{\rm H}(p,n){}^3{\rm He}$ cross section near $E_p=3.0$~MeV.
The data presented here will help to establish more accurately the properties
of these excited levels of ${}^4{\rm He}$.

\section{Thermonuclear Reaction Rate}
\label{sec:nasv}

The two-body thermonuclear reaction rate $N_A\langle\sigma v\rangle$ is
calculated from the cross section $\sigma$ using
\begin{equation}
N_A\langle\sigma v\rangle=\biggl({8\over\pi\mu}\biggr)^{1/2}
{N_A\over(kT)^{3/2}}\int_0^\infty E\,\sigma(E)\exp\biggl(-{E\over kT}\biggr)
\, dE~, \label{eq:nasv}
\end{equation}
where $N_A$ is Avogadro's number, $\mu$ is the reduced mass in the
entrance channel, $k$ is Boltzmann's constant, $T$ is temperature,
and $E$ is the center-of-mass energy.

Several tests were carried out to determine the effect of
the ${}^3{\rm He}(n,p){}^3{\rm H}$ reaction rate on
the primordial nucleosynthesis yields of the light elements
${}^2{\rm H}$, ${}^3{\rm He}$, ${}^4{\rm He}$, and ${}^7{\rm Li}$.
Standard big-bang nucleosynthesis calculations were performed
using the computer code described in Ref.~\cite{Kaw92}.
The calculation assumes that the baryon density is homogeneous,
and that there are three neutrino species.
We find that $\pm 10\%$ changes in the ${}^3{\rm He}(n,p){}^3{\rm H}$
reaction rate lead to changes of order 10\% in the ${}^3{\rm He}$ and
${}^7{\rm Li}$ abundances, changes of order 1\% in the ${}^2{\rm H}$
abundance, and no change in the ${}^4{\rm He}$ abundance.
The magnitude and direction of the changes are dependent upon the
value of the baryon-to-photon ratio $\eta$.
These findings are in agreement with the results of Smith, Kawano,
and Malaney~\cite{Smi93}.
In order to determine the energy range where the cross section
is important, we have also varied the reaction rate at different temperatures.
The final abundances of the light elements are found to depend
on the reaction rate in the temperature range $20 \le kT \le 60$~keV,
or equivalently $0.2\le T_9\le0.7$, where $T_9$ is the temperature in GK.
Changes in the reaction rate outside
of this temperature range do not affect the final abundances
(at least for baryon-to-photon ratios in the generally accepted range
$1\le 10^{10}\eta\le 10$).
This temperature range corresponds approximately to the
center-of-mass energy range $1\lesssim E\lesssim 250$~keV
in the $n-{}^3{\rm He}$ system, which is almost entirely covered by
the present experimental results.

\begin{figure}
\begin{center}
\includegraphics[bbllx=38,bblly=24,bburx=530,bbury=740,%
angle=90,width=3.4in]{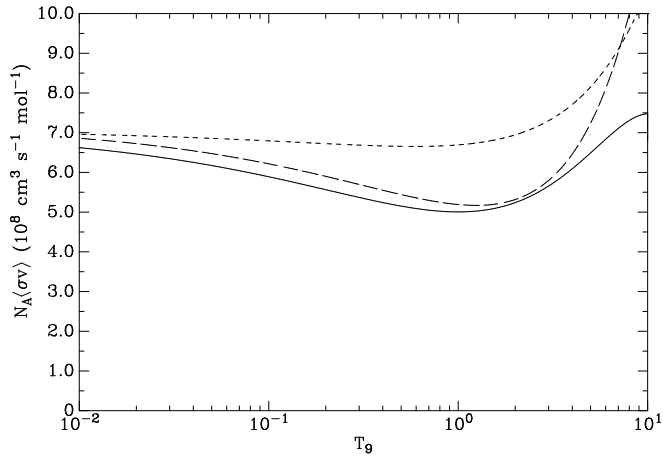}
\end{center}
\caption{The thermonuclear reaction rate $N_A\langle\sigma v\rangle$ for
${}^3{\rm He}(n,p){}^3{\rm H}$. The present determination is
given by the solid line; the rate from Smith, Kawano, and
Malaney~\protect\cite{Smi93} is given by the long-dashed line and the
rate from Caughlan and Fowler~\protect\cite{Cau88} by the short-dashed line.}
\label{fig:rate}
\end{figure}

For the calculation of the ${}^3{\rm He}(n,p){}^3{\rm H}$
thermonuclear reaction rate, we assumed the cross section
given by ENDF/B-VI evaluation~\cite{endf}.
This evaluation reproduces the well-known thermal cross section~\cite{Als64},
and as can be seen in Fig.~\ref{fig:snl} lies $\approx 3$\% higher
than the present experimental results for $E_n\le1$~MeV.
This deviation is considerably smaller than
our estimated systematic uncertainty in the experimental data.
Using this parameterization, the reaction rate
$N_A\langle\sigma v\rangle$ was then calculated by
numerically integrating Eq.~(\ref{eq:nasv}).
Our numerically integrated reaction rate is given within 1.5\%
for $T_9\le10$ by the following expression
(plotted in Fig.~\ref{fig:rate}):
\begin{equation}
  N_A\langle\sigma v\rangle=7.05\times 10^8(1-0.648T_9^{1/2}
  +0.426T_9-0.068T_9^{3/2}).
\end{equation}
We estimate the uncertainty in this reaction rate to be 5\% in the
temperature range important for big-bang nucleosynthesis calculations.
In the important temperature range,
the new rate is $\approx 5$\% lower than that given by
Smith, Kawano, and Malaney~\cite{Smi93}, and
15-25\% lower than that given by Caughlan and Fowler~\cite{Cau88}.

\section{Conclusions}

We have measured the ${}^3{\rm H}(p,n){}^3{\rm He}$ cross section
for $1.02\le E_p \le 4.50$~MeV, with a systematic uncertainty
estimated to be $5\%$. These measurements are considerably more accurate
than the previously available data over most of the energy range.
The data near $E_p=1.6$~MeV show the subtle effects of the $0^-$
second excited state of ${}^4{\rm He}$
which were not apparent in previous ${}^3{\rm H}(p,n){}^3{\rm He}$
cross section measurements.
In the future it may also be possible to compare the present
data to calculations which utilize realistic nuclear forces.
Calculations of this type have recently been performed for
$n-{}^3{\rm H}$ and $p-{}^3{\rm He}$ scattering at zero energy~\cite{Viv98}.

A new thermonuclear reaction rate for ${}^3{\rm He}(n,p){}^3{\rm H}$
has been calculated from the
ENDF/B-VI evaluation~\cite{endf} which agrees very
well with the present data and the
accurately known thermal cross section~\cite{Als64}.
This rate will allow for more accurate calculations of the
big-bang yields of ${}^3{\rm He}$ and ${}^7{\rm Li}$.
These calculations can then in turn be used to
test the consistency of the big-bang model as well as to
determine the baryon-to-photon ratio.

\acknowledgements

We thank G.~M.~Hale and M.~Drosg for providing
useful information concerning
the ${}^3{\rm H}(p,n){}^3{\rm He}$ reaction.
This work was supported in part by the National Science Foundation,
Grant No. PHY91-15574.
The final stages of this work were completed at the Triangle Universities
Nuclear Laboratory, supported in part by the U.~S.~Department of Energy,
Grant No.~DE-FG02-97ER41041.

\appendix
\section*{Previous Data}

As a result of our literature search for previous data, several
important aspects related to previous measurements have become apparent
which we would like to make more widely known.
This information should be particular useful to future evaluators
of the ${}^3{\rm H}(p,n){}^3{\rm He}$ or ${}^3{\rm He}(n,p){}^3{\rm H}$
cross sections. These aspects have not always been noted in the past.
The work of Drosg~\cite{Dro80} provided the basis for much of
this discussion.

(1) In addition to the data cited in the
Introduction, the ${}^3{\rm H}(p,n){}^3{\rm He}$ total cross section can
be inferred from other types of measurements.
For example, absolute $0^\circ$ differential cross section measurements
can be combined with angular distribution data to give total cross sections.
Data of this type are given in Refs.~\cite{Lis73,Dro80,Dro85,Dro87},
and references therein.
The total cross section for ${}^3{\rm He}(n,p){}^3{\rm H}$ can also be found
by subtracting the elastic neutron cross section from the
total neutron cross section (see Seagrave, Cranberg, and Simmons~\cite{Sea60}
and Alfimenkov {\em et al.}~\cite{Alf81}).
However, since the $(n,p)$ cross section is a small fraction of the
total cross section (for $E_n\gtrsim 50$~keV), this subtraction
can be subject to rather large uncertainties.

(2) Some ${}^3{\rm He}(n,p){}^3{\rm H}$ measurements are dependent
on the absolute $n-{}^3{\rm He}$ total cross section. There are significant
discrepancies in the existing $n-{}^3{\rm He}$ total cross section
measurements (see Ref.~\cite{Dro80} for details).
The $n-{}^3{\rm He}$ total cross section affects the data from
Refs.~\cite{Sea60,Alf81} mentioned in the previous paragraph
and also from Refs.~\cite{Say61,Cos70} which measured
the ratio of the ${}^3{\rm He}(n,p){}^3{\rm H}$ cross section
to the $n-{}^3{\rm He}$ total cross section.

(3) Some reported measurements have been superseded by or included
in subsequent data sets, and should be considered accordingly.
The data of Jarvis {\em et al.}~\cite{Jar50}
are renormalized and included with additional data in Ref.~\cite{la2014}.
The data of Ref.~\cite{la2014} are then apparently included in
revised form in Ref.~\cite{Per59} (see Ref.~\cite{Dro80}).
Also, the data of Macklin and Gibbons~\cite{Mac58} have been renormalized
in their later work~\cite{Gib59}.

\tighten

\begin{table}
\caption{Systematic errors in the absolute cross-section data
not included in the plotted error bars.
The total is computed by adding the individual errors in quadrature.}
\label{tab:errors}
\begin{tabular}{lc}
Source of Error                         & Error (\%) \\
\tableline
${}^3{\rm H}$ areal density             & 4 \\
neutron detection efficiency            & 3 \\
current integration                     & 1 \\
\tableline
total                                   & 5 \\
\end{tabular}
\end{table}

\end{document}